\documentclass[10pt]{article}

\usepackage[margin=1in]{geometry}
\usepackage[square,numbers,sort&compress]{natbib}
\usepackage[draft]{hyperref}
\usepackage{amsmath,amssymb,bm,mathtools}
\usepackage[T1]{fontenc}
\usepackage{lmodern,microtype}
\usepackage{titlesec}
\usepackage{graphicx}

\titleformat{\section}
  {\normalfont\large\bfseries}
  {\thesection.}{1em}{}

\titleformat{\subsection}
  {\normalfont\normalsize\bfseries}
  {\thesubsection.}{1em}{}
 
\begin{document}

\begin{center}
{\normalsize\bfseries
Thermodynamic Structure of Electronic Timekeeping\par}

\vspace{0.8em} 

{\normalsize Donhee Ham\par}

\vspace{0.3em}

{\small
John A. Paulson School of Engineering and Applied Sciences,\\
Harvard University, Cambridge, MA, USA\par}
\end{center}

\vspace{1em}

\begin{abstract}
\noindent Self-sustained electronic oscillators combining resonators and active circuits are the timekeepers of modern electronic systems, providing the temporal order essential to computing and communication. An oscillator can sustain periodic motion indefinitely. But it can never retain its initial timing information indefinitely. Thermal noise induces phase diffusion, causing an oscillator to gradually lose its initial timing information, typically over millions or billions of oscillation cycles or far more. A more precise clock, one whose phase diffuses more slowly, delays the inevitable loss of information for longer. Here we explore the thermodynamic structure underlying electronic timekeeping. An oscillator must continually increase the disorder of its environment to delay the loss of timing information. How much more disorder must it create to hold onto the same amount of timing information for longer? And how does the amount of disorder created compare with the amount of timing information the oscillator struggles to retain, only to eventually lose? The thermodynamic picture becomes richer when a less precise oscillator acquires and maintains relative-phase information with a more precise one and thus follows the more slowly diffusing phase of the latter, thereby delaying the loss of its own timing information. This reshaping of the information content of the oscillator pair must, in turn, reshape its thermodynamic structure. It is this thermodynamic interplay between the generation and transfer of temporal order, the loss of timing information, the acquisition and maintenance of relative-phase information, and the production of environmental disorder that we explore in this paper.
\end{abstract}

\section{\normalsize Introduction}

\noindent Resonator-based self-sustained oscillators are at the heart of modern electronic systems. Combining resonators with active transistor circuits that compensate for resonator losses, they autonomously generate periodic signals, providing the temporal order essential to digital computation, wireless and wireline communication, and a broad range of measurement and data acquisition systems \cite{SedraBook, HorowitzBook, LeeBook, RazaviBook, Armstrong, Ham2001, Andress2005}. Their fundamental role is to keep time. Yet an oscillator can never keep time perfectly. Thermal noise induces phase diffusion, causing an oscillator to gradually lose its initial timing information. A more precise clock is one whose phase diffuses more slowly, thereby delaying the inevitable loss of timing information for longer.

Underlying this electronic timekeeping is a rich thermodynamic structure. How much entropy must an oscillator dump into the environment to maintain a given timekeeping precision---or to delay the loss of timing information? How does this entropy cost compare with the amount of timing information the oscillator struggles to retain, only to eventually lose? Does the resonator's ability to retain energy, characterized by its quality factor, impart a distinctive thermodynamic signature to this cost? When a less precise oscillator is phase-locked to a more precise one---a common practice in electronics \cite{Adler, Viterbi, LeeBook, RazaviBook, PLL1, PLL2, PLL3}---the former gains relative-phase information with the latter and follows its phase, thereby delaying the loss of its timing information for longer. What is the thermodynamic cost associated with this precision transfer, which reshapes the information content of the oscillator pair? These are the questions we pursue in this paper. In this exploration, we draw on the methods of stochastic thermodynamics developed by Sekimoto, Hatano, Sasa, Seifert, and others \cite{Sekimoto,Hatano,Seifert0,Seifert}, particularly its systematic accounting of system entropy and environmental entropy production in nonequilibrium systems.

The thermodynamics of timekeeping per se has been studied in a variety of clock systems, including Brownian clocks, stand-alone damped resonators, resonator-based self-sustained mechanical oscillators, and resonator-less self-sustained electronic oscillators, often in regimes where the characteristic energy scales are comparable to or not far above the thermal energy \cite{Barato, Pietzonka, Gopal, Pearson, Izumida2026, Santolin2025, Gopal2026}. Our resonator-based self-sustained electronic oscillators exhibit distinctly different dynamics: their periodic motion is autonomously sustained against energy loss without external driving; the energy stored in their periodic motion exceeds the thermal energy by many orders of magnitude, often by seven to nine; and their phase diffusion occurs on a time scale many orders of magnitude longer than the oscillation period, typically spanning millions or billions of oscillation cycles or far more. These distinctive dynamics open up a broader set of thermodynamic questions, such as those posed above. In fact, previous studies have focused primarily on the first of these questions, arriving at various precision--cost relations, while the remaining three have largely been either inapplicable or left unexplored. It is the broader thermodynamic structure underlying the remarkably energetic, persistent, and highly coherent yet imperfect periodic motion of practical self-sustained electronic oscillators that we seek to uncover.

\section{\normalsize Coherence generation} \label{CG}

\begin{figure*}[t]
\centering
\includegraphics[width=0.8\textwidth]{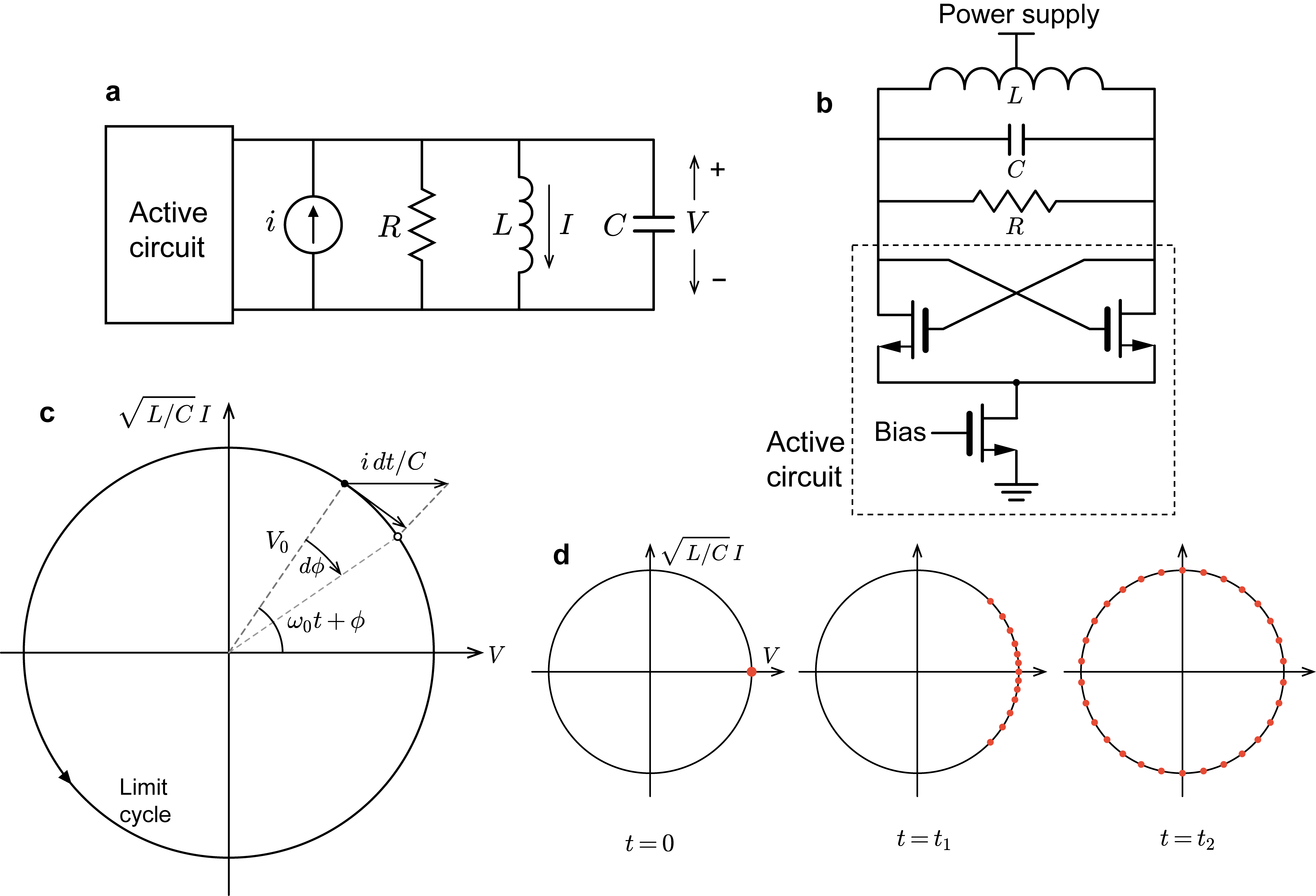}
\caption{\small
(a) A resonator-based self-sustained oscillator model. (b) A practical oscillator implementation. The active circuit is implemented using cross-coupled $n$MOS transistors. (c) Limit cycle in the $(V,\sqrt{L/C}I)$ phase space, where $V$ is the tank voltage and $I$ is the inductor current. Phase perturbation by noise $i(t)$ is also shown. (d) To visualize phase diffusion, we consider an ensemble of identical oscillators. It portrays the evolution of the phase probability density. At $t=0$, all oscillators have the same phase. As time progresses, their phases spread along the limit cycle. }
\label{fig:oscillator}
\end{figure*}

A resonator-based self-sustained oscillator consists of a resonator and an active circuit that compensates for the resonator's unavoidable parasitic loss \cite{SedraBook, HorowitzBook, LeeBook, RazaviBook,  Ham2001, Andress2005}. Figure \ref{fig:oscillator}a illustrates an oscillator model in which the resonator is a parallel LC tank, with resonant frequency \(\omega_0=(LC)^{-1/2}\), parasitic loss represented by a parallel resistance $R$, damping rate \(\gamma=1/(RC)\), and quality factor \(Q=\omega_0/\gamma\). With the active circuit compensating for the resonator's dissipative loss in $R$, the tank voltage \(V\) obeys a van der Pol-type stochastic differential equation \cite{Strogatz,Andronov},
\begin{equation}
\ddot V+[\gamma+g(V)]\dot V+\omega_0^2 V
=\frac{\dot i}{C}
\label{vdp}
\end{equation}
where \(i(t)\) is the noise current from both the resistor and the active circuit. The active circuit introduces \(g(V)\), whose form depends on the circuit topology. At small \(|V|\), \(g(V)\) is sufficiently negative that \(\gamma+g(V)<0\), causing oscillations to grow. At large \(|V|\), \(g(V)\) increases such that \(\gamma+g(V)>0\), causing oscillations to decay. The oscillator therefore self-starts from small fluctuations and self-sustains periodic oscillation in steady state. We do not seek a solution of Eq. (\ref{vdp}); we present it to identify the oscillator class.

\subsection{Timekeeping precision} 

For a self-sustained oscillator whose resonator has a sufficient quality factor, the steady-state tank voltage can be approximated as $V(t)=V_0\cos[\omega_0t+\phi(t)]$, where $V_0$ is the amplitude. This corresponds to a circular limit cycle in phase space (Fig. \ref{fig:oscillator}c). Here we have neglected amplitude fluctuations induced by the noise $i(t)$, since the limit cycle restores the amplitude toward $V_0$. In contrast, we have retained phase fluctuations $\phi(t)$ because the phase has no restoring mechanism and therefore undergoes diffusion \cite{Ham2003,Lax}. To determine the phase diffusion constant $D$, we first consider only the resistor's Nyquist noise, with $\langle i(t)i(t')\rangle=(2k_{\mathrm B}T/R)\delta(t-t')$, ignoring the active circuit noise, which we will account for shortly. At time $t$, when the total phase is $\omega_0t+\phi(t)$, the noise current $i(t)$ deposits charge $i(t)\,dt$ onto the capacitor over an infinitesimal interval $dt$, causing a voltage perturbation $i(t)\,dt/C$. Its projection onto the phase direction along the limit cycle is $V_0\,d\phi$, where $d\phi$ is the corresponding phase perturbation (Fig.~\ref{fig:oscillator}c). This leads to the Langevin equation for the phase \cite{Ham2003,Lax}
\begin{equation}
\dot\phi
=-\frac{i}{CV_0}\sin[\omega_0t+\phi].
\label{phaselangevin}
\end{equation}
In practical electronic oscillators, phase diffusion occurs on a time scale many orders of magnitude longer than the oscillation period. Thus, $\phi(t)$ varies negligibly over an oscillation period, and the period average of $\sin^2[\omega_0t+\phi(t)]$ is $1/2$  \cite{Lax}. With this averaging, the Langevin equation can be expressed in It\^o form as
\begin{equation}
d\phi=\sqrt{2D}\,dW,
\label{eq:phase}
\end{equation}
where $W$ is a Wiener process and the phase diffusion constant $D$ is given by \cite{Ham2003}
\begin{equation}
D=\frac{k_\mathrm{B}T}{E_0}\cdot\frac{\omega_0}{4Q}.
\label{Dideal}
\end{equation}
Here $E_0=CV_0^2/2$ is the energy stored in the resonator.

$D$ is the oscillator's timekeeping precision. In practical oscillators, phase diffusion occurs remarkably slowly compared with the oscillation itself. For example, a $1\,\mathrm{GHz}$ oscillator integrated on a silicon chip typically has $D$ on the order of $1\,\mathrm{s}^{-1}$ \cite{Ham2003}. This enormous separation of scales is hardly exceptional by engineering standards. In the frequency domain, the timekeeping precision appears as the spectral linewidth: the power spectrum of $V(t)=V_0\cos[\omega_0t+\phi(t)]$ is Lorentzian, with a linewidth of $2D$. Since $\omega_0/Q$ is the resonator linewidth and $E_0\gg k_{\mathrm B}T$ in Eq.~(\ref{Dideal}), the oscillator linewidth is dramatically compressed relative to the resonator linewidth. This narrow linewidth is the frequency-domain counterpart of phase diffusion occurring on a time scale far longer than the oscillation period.

We now express Eq.~(\ref{Dideal}) in terms of the power dissipation $P$ of the oscillator, while accounting for the active circuit noise. If a fraction $\eta$ of $P$ is dissipated in the resonator resistor $R$, with the remainder consumed by the active circuit, $\eta P=\omega_0E_0/Q$. We can also account for active circuit noise by replacing $k_{\mathrm B}T$ with $(1+\alpha)k_{\mathrm B}T$, where $\alpha$ represents the active circuit noise relative to the resistor thermal noise in the resonator. Equation~(\ref{Dideal}) then becomes $D=c\,({k_{\mathrm B}T}/{P}) ({\omega_0^2}/{4Q^2})$, where $c=(1+\alpha)/\eta$ \cite{Ham2003}. The factor $c$ can be further modified to account for deviations from a sinusoidal waveform \cite{Demir,HL,Kundert}. For the remainder of this paper, we adopt the minimal model with $c=1$---i.e., $\alpha = 0$ (no active circuit noise), $\eta = 1$ (all power dissipation in the resonator), and sinusoidal oscillation---because $c$ is typically of order unity and, more importantly, does not alter the essential thermodynamic picture. We thus use \cite{Ham2003}:
\begin{equation}
D=\frac{k_{\mathrm B}T}{P}\frac{\omega_0^2}{4Q^2}.
\label{eq:D}
\end{equation}
The scaling $D\propto T/(PQ^2)$ is consistent with Leeson's classic frequency-domain phase-noise formula~\cite{Leeson}, although Leeson's formula does not explicitly invoke the phase diffusion coefficient $D$. In the following subsections, we consider the thermodynamics of a stand-alone oscillator, examining how its timekeeping precision $D$ relates to the oscillator entropy $S_1$ and the environmental entropy $S_2$. 

\subsection{Oscillator entropy $S_1$} \label{firstS1}

As its phase diffuses, an oscillator gradually loses its initial timing information (Fig.~\ref{fig:oscillator}d). In a more precise clock, phase diffuses more slowly (smaller $D$), thus postponing the inevitable loss of timing information. The oscillator entropy, given by the Shannon--Gibbs expression \cite{Hatano, Seifert0} $S_1=-k_{\mathrm B}\int p(\phi,t)\ln p(\phi,t)\,d\phi$, quantifies this timing information loss, where $p(\phi,t)$ is the probability density of $\phi$ corresponding to the phase diffusion. 

Let phase diffusion start from a sharply defined initial phase. In the unwrapped phase representation, $-\infty<\phi<\infty$,  the Gaussian probability density $p(\phi,t)=(4\pi Dt)^{-1/2}\exp[-\phi^2/(4Dt)]$ describes phase diffusion with $\langle \phi^2(t)\rangle=2Dt$. Then $S_1=({k_{\mathrm B}}/{2})\ln(4\pi eDt)$ and $\dot S_1(t)={k_{\mathrm B}}/({2t})$. $S_1$ grows monotonically without bound, but at an ever-decreasing rate $\dot S_1$ that asymptotically vanishes. In the wrapped phase representation, $-\pi\leq\phi<\pi$, $p(\phi,t)=(1/2\pi)[1+2\sum_{n=1}^{\infty}e^{-n^2Dt}\cos(n\phi)]$. Then $S_1$ (for which no simple closed-form expression is available) increases monotonically toward $k_{\mathrm B}\ln(2\pi)$ as the phase distribution approaches the uniform distribution, but again at an ever-decreasing rate $\dot S_1$ that asymptotically vanishes. The limiting expressions illustrate this behavior: for $Dt\ll1$, $S_1(t)\simeq({k_{\mathrm B}}/{2})\ln(4\pi eDt)$ and $\dot S_1(t)\simeq{k_{\mathrm B}}/({2t})$, just like in the unwrapped representation; for $Dt\gg1$, $S_1(t)\simeq k_{\mathrm B}\ln(2\pi)-k_{\mathrm B}e^{-2Dt}$, while $\dot S_1(t)\simeq2k_{\mathrm B}De^{-2Dt}$.

Whether in the wrapped or unwrapped representation, the monotonically increasing entropy $S_1$ reflects the accumulating loss of timing information. Its ever-decreasing rate $\dot S_1$, with $\dot S_1 \rightarrow 0$, means that the oscillator loses progressively less timing information per unit time. Since $p(\phi,t)$ depends only on the product $Dt$, whether wrapped or unwrapped, reducing $D$ to $D/N$ prolongs the time required for the same increase in $S_1$---the same loss of timing information---by a factor of $N$, starting from the same phase probability distribution. Thus, a more precise clock delays the inevitable loss of timing information for proportionally longer.

How much timing information does a practical oscillator lose over the characteristic phase diffusion time $1/D$? Consider a 1-GHz oscillator ($T_0=2\pi/\omega_0=1$ ns) with $D=1\,\mathrm{s}^{-1}$ and a perfectly known phase at $t=0$. Over the interval from $t=T_0$ to $t=T_0+1/D$---choosing $t=T_0$ rather than $t=0$ avoids the singularity $S_1(0)=-\infty$ arising from the idealized assumption of a perfectly known initial phase---$\Delta S_1\approx10.4\,k_{\mathrm B}$ and $10.3\,k_{\mathrm B}$ in the unwrapped and wrapped representations, respectively. Thus, over the time $1/D=1$ s, the oscillator loses approximately 15 bits of timing information. If $D$ is reduced by a factor of 10, the same amount of information ($\sim15$ bits) is lost over a tenfold longer duration (10 s), starting from $t=10T_0$ to ensure the same initial phase distribution.

\subsection{Environmental entropy production $S_2$} \label{firstS2}

While the oscillator entropy $S_1$ measures the accumulated loss of timing information for a given timekeeping precision $D$, the increase in the environmental entropy $S_2$, or the environmental entropy production, measures the thermodynamic cost of maintaining that precision $D$. We can see this by substituting \(\dot S_2=P/T\) \cite{Sekimoto,Hatano,Seifert0,Seifert} into Eq. (\ref{eq:D}): 
\begin{equation}
D\dot S_2
=\frac{k_{\mathrm B}\omega_0^2}{4Q^2}.
\label{eq:main}
\end{equation}
The oscillator pays for maintaining its precision $D$ by dumping entropy at the constant rate of $\dot S_2 > 0$ into the environment, continually increasing its disorder; this contrasts with the fact that the oscillator accumulates increasingly less disorder per unit time given $\dot S_1\rightarrow0$. For a given resonator quality factor $Q$, a smaller $D$---more precise timekeeping---requires a higher environmental entropy production rate. In fact, $Q$ fixes the exchange rate between clock precision and thermodynamic cost: for a given clock precision, the thermodynamic cost decreases as $1/Q^2$. This reveals that the resonator's ability to retain energy imparts a distinctive thermodynamic signature to electronic timekeeping. Equation~(\ref{eq:main}) can be written as  
\begin{equation}
\overline{\sigma^2}\,{\bar s_2 }
=\frac{2\pi^2 }{Q^2},
\label{eq:cycle}
\end{equation}
where $\overline{\sigma^2}\equiv2DT_0$ is the per-period phase variance accumulation, and $\bar s_2\equiv\dot S_2T_0 / k_{\mathrm B}$ is the per-period environmental entropy production in units of $k_{\mathrm B}$. This expression makes the role of $Q$ particularly transparent. As the derivation of Eq. (\ref{eq:main}) from Eq. (\ref{eq:D}) above makes clear, the precision-cost relation of Eq. (\ref{eq:main}) has long been implicit in the expression for $D$ [Eq. (\ref{eq:D})] in resonator-based self-sustained electronic oscillators \cite{Leeson,Ham2003}, and is made explicit here through the systematic entropy accounting of stochastic thermodynamics \cite{Sekimoto,Hatano,Seifert0,Seifert}.

The thermodynamic cost of timekeeping has also been studied in various other clock systems \cite{Barato, Pietzonka, Gopal, Pearson, Izumida2026, Santolin2025, Gopal2026}, yielding various results for $\overline{\sigma^2}\bar s_2$. Some of these results fall within a framework called the thermodynamic uncertainty relation (TUR), which, for example, gives $\overline{\sigma^2}\bar s_2\geq8\pi^2$ \cite{Barato}. $\overline{\sigma^2}\bar s_2$ for our practical resonator-based self-sustained oscillator takes the distinct form of Eq.~(\ref{eq:cycle}), with $Q$ setting the exchange rate between clock precision and thermodynamic cost. For high $Q$, this product can be far smaller than $8\pi^2$. The underlying principle remains the same: greater precision requires greater entropy production.

\subsection{Timing information is cheap; slowing its loss is expensive} \label{cheapexpensive}

$S_1$ describes the physics of timing information loss. $S_2$ describes the physics of the cost of delaying that information loss. To maintain its timekeeping precision $D$, the oscillator pays a thermodynamic cost by dumping entropy into the environment at a constant rate $\dot S_2>0$, even as its own entropy $S_1$ continues to grow due to a finite $D$. Thus, despite paying continuously, the oscillator gradually loses its initial timing information. 

Improving clock precision by a factor of $N$, reducing $D$ to $D/N$, prolongs the time required to lose the same amount of information by a factor of $N$, as seen in Sec.~\ref{firstS1}, but at a greater total environmental entropy cost. If the reduction to $D/N$ is achieved by increasing power $P$ by a factor of $N$ at fixed $Q$, $\dot S_2$ increases by a factor of $N$ [Eqs.~(\ref{eq:D}) and (\ref{eq:main})], and the environmental entropy produced over the $N$-times longer interval increases by a factor of $N^2$. If the reduction to $D/N$ is achieved by increasing $Q$ by a factor of $\sqrt{N}$ at fixed $P$, $\dot S_2$ remains unchanged [Eqs.~(\ref{eq:D}) and (\ref{eq:main})], and the environmental entropy produced over the $N$-times longer interval increases by a factor of $N$. Either way, delaying information loss comes at a greater thermodynamic cost. The oscillator does not pay to preserve its timing information indefinitely; it pays to slow down its inevitable loss.

How large is this thermodynamic cost in a practical oscillator, compared with the amount of timing information it struggles to retain, only to eventually lose? In the 1-GHz oscillator example ($T_0=1$ ns; $D=1\,\mathrm{s}^{-1}$) of Sec.~\ref{firstS1}, the oscillator loses approximately 15 bits of timing information over the characteristic phase diffusion time $1/D$, from $t=T_0$ to $t=T_0+1/D$. The environmental entropy produced over the same interval follows from Eqs.~(\ref{eq:main}) and (\ref{Dideal}): $\Delta S_2=\dot S_2/D=4k_{\mathrm B}(E_0/k_{\mathrm B}T)^2$. For a typical stored energy of $E_0=1\,\mathrm{pJ}$ at room temperature, the oscillator dumps approximately $\Delta S_2\approx2.4\times10^{17}k_{\mathrm B}$ of entropy into the environment, equivalent to $\sim3.5\times10^{17}$ bits. The contrast is striking: the oscillator struggles to retain $\sim15$ bits of timing information by paying an enormous thermodynamic cost equivalent to $\sim3.5\times10^{17}$ bits, only to lose that information.

The physical origin of this enormous thermodynamic cost becomes apparent if we remove the active circuit [i.e., set $g(V)=0$ in Eq.~(\ref{vdp})] at $t=T_0$ in the 1-GHz oscillator example above and let the resonator ring down. Neglecting amplitude fluctuations, the phase diffusion constant in the damped resonator can be shown to take the same form as in Eq.~(\ref{Dideal}), with the decaying resonator energy $E(t)=E_0e^{-\gamma(t-T_0)}$ replacing $E_0$. The phase diffusion constant then grows as $D(t)=De^{\gamma(t-T_0)}$, accelerating the loss of timing information. The loss of $\sim15$ bits of timing information that takes $1/D=1$~s (from $t=T_0$ to $t=T_0+1/D$) in the self-sustained oscillator occurs in as little as $\sim\gamma^{-1}\ln(4E_0/k_{\mathrm B}T)\sim20$~ns in the damped resonator. Over this $\sim20$~ns interval, the resonator dissipates virtually all of its initial energy $E_0$ (since $1/\gamma\sim1$~ns), after which there is no net heat release. The environmental entropy produced by the damped resonator is thus $\Delta S_{2,\mathrm{damped}}\approx E_0/T\sim2.4\times10^8k_{\mathrm B}$, which is remarkably smaller than $\Delta S_2 \sim2.4\times10^{17}k_{\mathrm B}$ produced over the 1-s interval in which the self-sustained oscillator loses the same $\sim 15$~bits of timing information. This striking contrast shows that the enormous thermodynamic cost of self-sustained timekeeping arises precisely because the stored energy $E_0$ must be continually replenished to keep $D$ small and thereby delay the loss of timing information.  

As an important aside, in the damped resonator, since $\dot S_2(t)=(-dE/dt)/T=(\gamma E_0/T)e^{-\gamma(t-T_0)}$, we have $D(t)\dot S_2(t)=k_{\mathrm B}\omega_0^2/(4Q^2)$, just as in the self-sustained oscillator, for which $D\dot S_2=k_{\mathrm B}\omega_0^2/(4Q^2)$ [Eq.~(\ref{eq:main})]. The form of the phase diffusion constant underlying these two products---Eq.~(\ref{Dideal}) for the self-sustained oscillator and Eq.~(\ref{Dideal}) with $E_0$ replaced by $E(t)$ for the damped resonator---assumes negligible thermal amplitude fluctuations relative to the amplitude itself. The product $D(t)\dot S_2(t)=k_{\mathrm B}\omega_0^2/(4Q^2)$ in the damped resonator is thus valid only while the decaying resonator amplitude remains sufficiently large (or, in terms of energy, $E(t)\gg k_{\mathrm B}T$). In contrast, the product $D\dot S_2=k_{\mathrm B}\omega_0^2/(4Q^2)$ in the self-sustained oscillator remains valid in steady state because the oscillator not only maintains a large amplitude but also suppresses amplitude fluctuations through limit-cycle restoration. This reveals that the precision--cost relation $D\dot S_2=k_{\mathrm B}\omega_0^2/(4Q^2)$ in the self-sustained oscillator [Eq.~(\ref{eq:main})] has its physical origin in the dynamics of an energetically excited resonator, yet it is the self-sustaining mechanism that enables the relation to remain valid indefinitely.

\section{\normalsize Coherence transfer} \label{CT}

\subsection{Physics of phase locking} 

\begin{figure*}[t]
\centering
\includegraphics[width=0.7\textwidth]{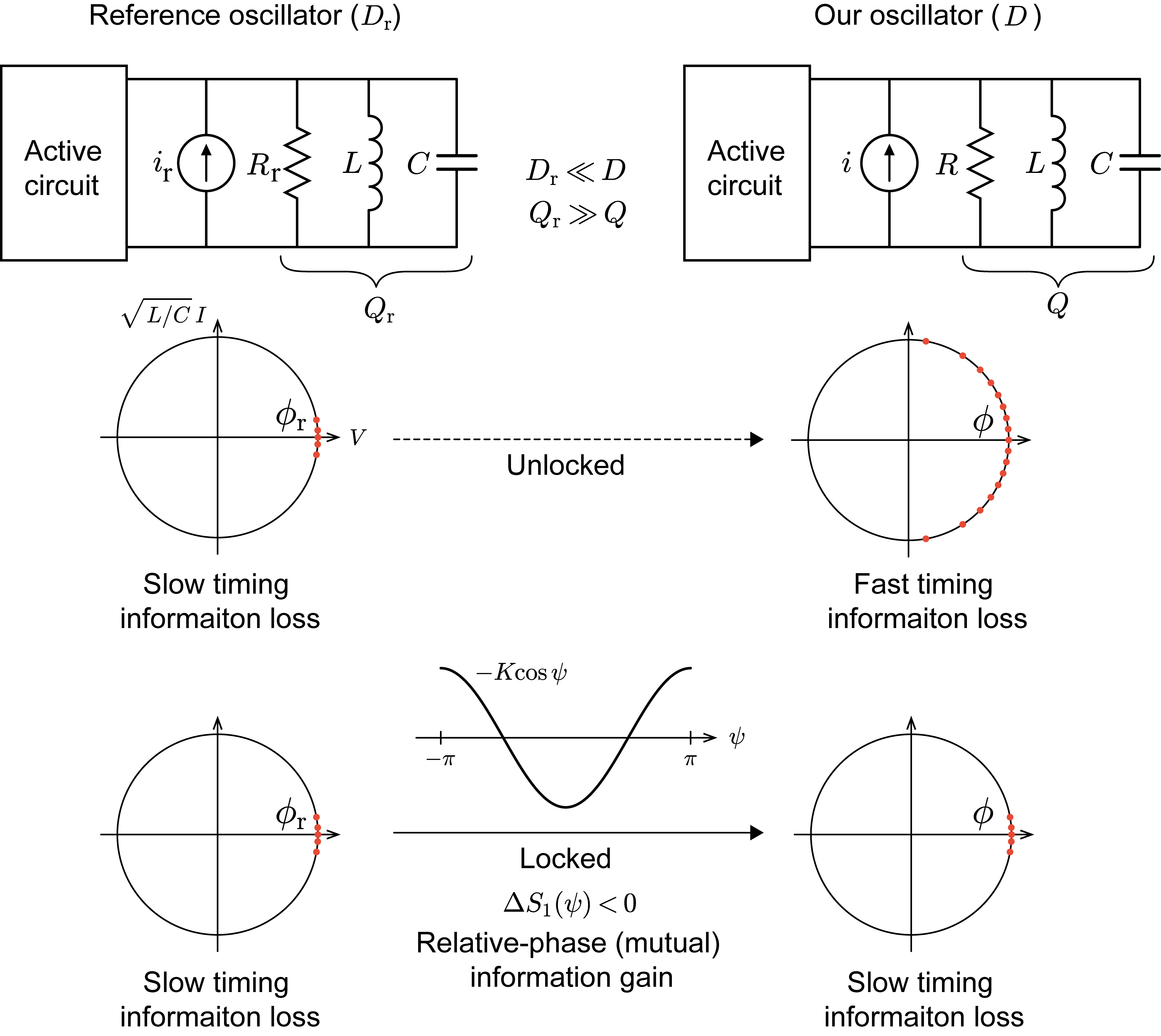}
\caption{\small Phase locking of our oscillator with a lower resonator quality factor $Q$ and lower timekeeping precision $D$ to a reference oscillator with a higher resonator quality factor $Q_\mathrm{r}$ and higher timekeeping precision $D_\mathrm{r}$. }
\label{fig:pl}
\end{figure*}

Consider two independent oscillators (Fig.~\ref{fig:pl}). Their phase diffusion dynamics are described by $d\phi=\sqrt{2D}\,dW$ and $d\phi_\mathrm{r}=\sqrt{2D_\mathrm{r}}\,dW_\mathrm{r}$, where $D$ and $D_\mathrm{r}$ are phase diffusion constants and $W$ and $W_\mathrm{r}$ are independent Wiener processes. Their resonator quality factors are $Q$ and $Q_\mathrm{r}$. We take the first to be the oscillator studied above (`our oscillator') and the second to be a much more precise oscillator (`reference'), with \(D_\mathrm{r}\ll D\) and \(Q_\mathrm{r}\gg Q\). 

We seek to phase lock our oscillator to the reference---widely employed in electronics via phase-locked loops (PLLs) and injection locking \cite{Adler, Viterbi, LeeBook, RazaviBook, PLL1, PLL2, PLL3}---to transfer the reference's superior timekeeping precision $D_\mathrm{r}$ to our oscillator. Phase locking also synchronizes frequencies, but we assume zero detuning to focus on precision transfer. Under locking, the reference phase $\phi_\mathrm{r}$ continues to diffuse according to $d\phi_\mathrm{r}=\sqrt{2D_\mathrm{r}}\,dW_\mathrm{r}$, while the relative phase $\psi=\phi-\phi_\mathrm{r}$---our oscillator's phase $\phi$ relative to the reference phase $\phi_\mathrm{r}$---follows \cite{Viterbi,Adler}
\begin{equation}
d\psi=-K\sin\psi\,dt+\sqrt{2D}\,dW.
\label{relativephase}
\end{equation}
$K$ is the locking strength, and the reference noise has been ignored given $D_\mathrm{r}\ll D$. $\psi$ does not diffuse freely due to the restoring term $-K\sin\psi$, which corresponds to a phase potential $-K\cos\psi$ and tends to localize $\phi$ around $\phi_\mathrm{r}$, or $\psi$ around $0$ in the wrapped representation. The steady-state probability density of $\psi$ in the wrapped representation is the von Mises distribution: 
\begin{equation}
p(\psi)=\frac{e^{a\cos\psi}}{2\pi I_0(a)},
\qquad a\equiv\frac{K}{D},
\label{vonmises}
\end{equation}
where $I_0$ is the modified Bessel function of the first kind of order zero. In typical GHz PLLs and oscillators, $K\sim10^4$--$10^7\,\mathrm{s}^{-1}$ and $D\sim1$--$10\,\mathrm{s}^{-1}$, giving $a\sim10^3$--$10^7$. In this practical strong-locking regime ($a \gg1$), $p(\psi)$ is sharply localized, with $\langle\psi^2\rangle\simeq 1/a \ll1$. Thus, $\phi$ closely follows $\phi_\mathrm{r}$, which continues to diffuse with constant \(D_\mathrm{r}\), and inherits the reference's timekeeping precision $D_\mathrm{r}$, rather than its own precision $D$ (Fig.~\ref{fig:pl}). In other words, by establishing relative-phase information with the reference, our oscillator follows the more slowly diffusing reference phase, thereby delaying the loss of its own timing information for longer. Phase locking thus reshapes the information content of the oscillator pair. How does this alter its thermodynamic structure?

\subsection{Phase entropies $S_1$} \label{infotheory}

The oscillator entropy $S_1$ introduced in Sec. \ref{firstS1} describes the Shannon--Gibbs entropy of our oscillator's phase $\phi$. We now additionally apply the same measure to $\phi_\mathrm{r}$, $\psi$, and the joint variable $(\phi$, $\phi_\mathrm{r})$. We denote these four entropies by $S_1(\phi)$, $S_1(\phi_\mathrm{r})$, $S_1(\psi)$, and $S_1(\phi, \phi_\mathrm{r})$, referring to them as phase entropies. We focus on the wrapped representation for simplicity. Whether locked or unlocked, all four phase entropies approach steady-state values (Table~\ref{tab:phase_entropy}), and thus their time derivatives eventually vanish, $\dot S_1\to0$. 

Whether locked or unlocked, $S_1(\phi)$ and $S_1(\phi_\mathrm{r})$ describe the loss of timing information in the respective oscillators. Both increase toward $k_{\mathrm B}\ln(2\pi)$. This is because both phases continue to diffuse---with $D$ and $D_\mathrm{r}$ when unlocked, and $D_\mathrm{r}$ for both when locked---and their wrapped phase distributions become uniform. $S_1(\psi)$ tells a richer story. When the oscillators run unlocked, $\phi$ and $\phi_\mathrm{r}$  diffuse, so $\psi$ diffuses toward a uniform distribution, increasing $S_1(\psi)$ toward $k_{\mathrm B}\ln(2\pi)$. Now turn on phase locking after this unlocked steady state is reached. Then $\psi$ becomes localized according to Eq.~(\ref{vonmises}), and its entropy is reduced toward a new steady-state value. The entropy change from the unlocked to the locked steady state is  
\begin{equation}
\Delta S_1(\psi)=k_{\mathrm B}\left[\ln I_0(a)-a\frac{I_1(a)}{I_0(a)}\right]<0,
\label{S1diff}
\end{equation}
where $I_1$ is the modified Bessel function of the first kind of order one. This reduction in relative-phase entropy can equivalently be understood as a reduction in joint phase entropy, since $S_1(\phi,\phi_\mathrm{r})=S_1(\phi_\mathrm{r})+S_1(\psi)=k_{\mathrm B}\ln(2\pi)+S_1(\psi)$ in both the unlocked and locked steady states. The locked pair with aligned phases is therefore more ordered than the unlocked pair. This entropy reduction reflects a gain in relative-phase information, which is precisely the mutual information $I(\phi;\phi_\mathrm{r})=-\Delta S_1(\psi)/k_{\mathrm B}$ between the two oscillators. For $a=10^3$ and $10^7$ in the practical strong-locking regime ($a\gg1$), $\Delta S_1(\psi)$ due to phase locking is approximately $-3.9k_{\mathrm B}$ and $-8.5k_{\mathrm B}$, respectively, corresponding to relative-phase (or equivalently, mutual) information gains of 5.6 and 12.2 bits.

\begin{table}[t]
\centering
\begingroup
\renewcommand{\arraystretch}{1.2}
\setlength{\tabcolsep}{12pt}
\begin{tabular}{ccc}
\hline
Entropy & Unlocked & Locked \\
\hline
$S_1(\phi)$
& $k_\mathrm{B}\ln(2\pi)$
& $k_\mathrm{B}\ln(2\pi)$ \\

$S_1(\phi_\mathrm{r})$
& $k_\mathrm{B}\ln(2\pi)$
& $k_\mathrm{B}\ln(2\pi)$ \\

$S_1(\psi)$
& $k_\mathrm{B}\ln(2\pi)$
& $k_\mathrm{B}\ln(2\pi)+k_\mathrm{B}\left[
\ln I_0(a)
-a {I_1(a)}/{I_0(a)}
\right]$ \\

$S_1(\phi,\phi_\mathrm{r})$
& $2k_\mathrm{B}\ln(2\pi)$
& $2k_\mathrm{B}\ln(2\pi)+k_\mathrm{B}\left[
\ln I_0(a)
-a{I_1(a)}/{I_0(a)}
\right]$ \\
\hline
\end{tabular}
\endgroup
\caption{\small Steady-state phase entropies in the oscillator pair without and with locking.}
\label{tab:phase_entropy}
\end{table}

\subsection{Environmental entropy production $S_2$} \label{phaselockS2}

The total environmental entropy production rate from the oscillator pair and its locking circuitry is
\begin{equation}
\dot S_{2,\mathrm{pair}}
=\frac{k_{\mathrm B}\omega_0^2}{4Q_\mathrm{r}^2D_\mathrm{r}}
+\frac{k_{\mathrm B}\omega_0^2}{4Q^2D}
+\frac{P_\mathrm{L}(t)}{T}.
\label{S2pair}
\end{equation}
The first two terms, obtained from Eq.~(\ref{eq:main}), account for the entropy produced by the power dissipation of the reference oscillator and our oscillator, respectively. Although phase locking transfers the reference's precision $D_\mathrm{r}$ to our oscillator, the latter's original power dissipation remains unchanged, hence the free-running $D$ in the second term. The third term, $P_\mathrm{L}(t)/T\equiv\dot S_{2,\mathrm{L}}$, arises from the power $P_\mathrm{L}(t)$ dissipated by the circuitry that implements the phase-locking dynamics of Eq.~(\ref{relativephase}) to establish and maintain phase alignment. In general, $P_\mathrm{L}(t)$ varies during the establishment of phase alignment and assumes a steady-state value $P_\mathrm{L}(\infty)$ to maintain phase alignment once locking is established. To establish phase alignment and thereby bring about the reduction in joint phase entropy $\Delta S_1(\psi)$ given in Eq.~(\ref{S1diff})---or equivalently, the gain in relative-phase information---the locking circuitry must increase the disorder of the environment. In the thermodynamic description of phase alignment, the associated environmental entropy produced, $\Delta S_{2,\mathrm{L}}=\int P_\mathrm{L}(t)/T\,dt$ over the locking transient, must at least compensate for the reduction in joint phase entropy, such that $\Delta S_{2,\mathrm{L}}+\Delta S_1(\psi)\geq0$. Equality corresponds to the reversible limit of phase alignment.

How much more thermodynamically economical is it to operate our oscillator at the higher timekeeping precision $D_\mathrm{r}$ once phase locking is established than to sustain the same precision autonomously by increasing its power dissipation at its given resonator quality factor $Q$? For the former, the environmental entropy production rate in the locked steady state is $\dot S_{2,\mathrm{pair}}$ in Eq.~(\ref{S2pair}) with $P_\mathrm{L}(t)=P_\mathrm{L}(\infty)$. For the latter, it is $\dot S_{2,\mathrm{solo}}=k_{\mathrm B}\omega_0^2/(4Q^2D_\mathrm{r})$, according to Eq.~(\ref{eq:main}). The ratio of the two entropy production rates is then
\begin{equation}
\frac{\dot S_{2,\mathrm{pair}}}{\dot S_{2,\mathrm{solo}}}
=\left(\frac{Q}{Q_\mathrm{r}}\right)^2
+(1+\beta)\frac{D_\mathrm{r}}{D}\ll1.
\label{pairvsolo}
\end{equation}
Here $\beta\equiv P_\mathrm{L}(\infty)/P$, where $P=k_{\mathrm B}T\omega_0^2/(4Q^2D)$ is our oscillator's original power dissipation corresponding to its free-running $D$. For $Q_\mathrm{r}\gg Q$, $D_\mathrm{r}\ll D$, and $\beta$ of order unity \cite{RazaviBook,LeeBook}, coherence transfer via phase locking achieves the higher timekeeping precision $D_\mathrm{r}$ at a substantially lower environmental entropy production rate than autonomous coherence generation. Importing precision from a reference oscillator with a higher-quality resonator is far less costly than generating the same precision by increasing the power dissipation of our oscillator with a lower-quality resonator.

In summary, through phase locking, our oscillator acquires and retains relative-phase information $I(\phi;\phi_\mathrm{r})=-\Delta S_1(\psi)/k_\mathrm{B}$. By retaining this information, our oscillator follows the more slowly diffusing reference phase and thereby slows the loss of its own timing information, as reflected in the slower evolution of $S_1(\phi)$. Remarkably, slowing the loss of timing information by retaining relative-phase information is thermodynamically far less costly [Eq. (\ref{pairvsolo})] than slowing the loss of timing information by autonomous coherence generation. Yet even this substantially reduced thermodynamic cost is enormous compared with both the amount of relative-phase information retained and the amount of timing information eventually lost. This is the thermodynamic structure associated with the reshaping of the information content of the oscillator pair through phase locking.

\subsection{The thermodynamic economics of timekeeping} 

\begin{table*}[t]
\centering

\caption{\small Timekeeping precision, information, and thermodynamic cost across oscillator configurations.}
\label{tab:cost_comparison}

\begingroup
\small

\renewcommand{\arraystretch}{1.5}
\setlength{\tabcolsep}{5pt}

\begin{tabular}{@{}lcccc@{}}

\hline
\noalign{\vskip 4pt}

\parbox[c]{4.5cm}{\raggedright }
& \parbox[c]{2.3cm}{\centering Our\\oscillator}
& \parbox[c]{2.3cm}{\centering Our oscillator,\\power increased}
& \parbox[c]{2.3cm}{\centering Our oscillator,\\phase locked}
& \parbox[c]{2.3cm}{\centering Reference\\oscillator} \\

\noalign{\vskip 4pt}
\hline
\noalign{\vskip 4pt}

Resonator
& $Q$ & $Q$ & $Q,\ \sqrt{N}Q$ & $\sqrt{N}Q$ \\

Power
& $P$ & $NP$ & $(2+\beta)P$ & $P$ \\

Precision
& $D$ & $D/N$ & $D/N$ & $D/N$ \\

Duration
& $1$ & $N$ & $N$ & $N$ \\

\noalign{\vskip 4pt}
\hline
\noalign{\vskip 4pt}

$\Delta S_1(\phi)$ (timing information loss)
& $1$ & $1$ & $1$ & $1$ \\

$\Delta S_1(\psi)$ (mutual information gain)
& --- & --- & $<0$ & --- \\

\noalign{\vskip 4pt}
\hline
\noalign{\vskip 4pt}

$\dot S_2$ (thermodynamic cost rate)
& $1$ & $N$ & $2+\beta$ & $1$ \\

$\Delta S_2$ (thermodynamic cost)
& $1$ & $N^2$ & $(2+\beta)N$ & $N$ \\

\noalign{\vskip 4pt}
\hline

\end{tabular}

\vspace{4pt}

\begin{minipage}{0.95\textwidth}
\footnotesize

Duration, $\Delta S_1(\phi)$, $\dot S_2$, and $\Delta S_2$ are normalized to their respective values for our original oscillator.
$\Delta S_1(\phi)$ measures the accumulated loss of timing information over the indicated duration during steady-state oscillation ($\phi_\mathrm{r}$ for the reference oscillator). $\Delta S_1(\psi)$ reflects the gain in relative-phase information
(mutual information) upon phase locking. For the phase-locked configuration, power dissipation and environmental
entropy production include contributions from both oscillators and the locking circuitry.

\end{minipage}

\endgroup
\end{table*}

The preceding analyses in this section as well as in Sec.~\ref{CG} have shown that greater timekeeping precision comes at a greater thermodynamic cost, but the precise relationship between the two depends on how that precision is achieved. To provide a unified view, we consolidate the precision-cost relations in Table~\ref{tab:cost_comparison} for a common factor $N \gg 1$ of precision improvement and a common amount of timing information lost, starting from the same initial phase distribution. Concretely, we compare four configurations: our normally operating oscillator with precision $D$, resonator quality factor $Q$, and power dissipation $P$; our oscillator with its power dissipation increased to $NP$ to achieve precision $D/N$; the reference oscillator with precision $D_\mathrm{r}=D/N$, achieved with $Q_\mathrm{r}=\sqrt{N}Q$ at the same power dissipation $P$ as our normally operating oscillator [Eq.~(\ref{eq:D})]; and our oscillator phase-locked to the reference to inherit precision $D/N$. Starting from the same initial phase distribution, the latter three all take $N$ times longer than our original oscillator to lose the same amount of timing information.

Relative to our original oscillator, all three higher-precision configurations pay the price for the better precision, incurring greater environmental entropy production to lose the same amount of information, $\Delta S_1(\phi)/k_\mathrm{B}$, but to markedly different degrees. Increasing our oscillator's power incurs an $N^2$-fold cost, whereas coherence transfer incurs a much lower $(2+\beta)N$-fold cost. The reference oscillator itself incurs the lowest cost, $N$ times that of our original oscillator. Why, then, not use the reference oscillator directly? A high-quality resonator can easily lose its high quality factor when coupled to external loads. Phase locking allows the reference to preserve its high resonator quality while our oscillator serves as the world-facing clock. At a modest additional thermodynamic cost, coherence transfer thus enables our oscillator to deliver the reference's superior timekeeping precision without compromising the reference resonator. This reveals a thermodynamic rationale for the widespread use of phase locking in practical electronics. Finally, however much the thermodynamic cost $\Delta S_2$ is reduced, in practice it still far exceeds $\Delta S_1(\phi)$, reflecting the timing information the oscillator strives to retain but eventually loses, and, where applicable, $|\Delta S_1(\psi)|$, reflecting the relative-phase information gained.

\subsection{Landauer limit of phase alignment} \label{sec:LL}

As discussed in Sec.~\ref{phaselockS2}, the circuitry that realizes the phase-locking dynamics of Eq.~(\ref{relativephase}) dissipates power $P_\mathrm{L}(t)$ to establish and maintain phase alignment. The third term in Eq.~(\ref{S2pair}), $P_\mathrm{L}(t)/T\equiv\dot S_{2,\mathrm{L}}$, is the associated environmental entropy production rate. Since establishing phase alignment---gaining relative-phase information---reduces joint phase entropy by $\Delta S_1(\psi)$ in Eq.~(\ref{S1diff}), the environmental entropy produced by the locking circuitry over the locking transient, $\Delta S_{2,\mathrm{L}}=\int P_\mathrm{L}(t)/T\,dt$, must satisfy $\Delta S_{2,\mathrm{L}}+\Delta S_1(\psi)\geq0$. In other words, the decrease in the disorder of the oscillator pair must be accompanied by an equal or greater increase in environmental disorder created by the locking circuitry. The equality, $\Delta S_{2,\mathrm{L}}=-\Delta S_1(\psi)$, defines the reversible limit of phase alignment \cite{Hatano, Seifert0}. Although this limit is far out of reach in practice, we show that it can be attained in an idealized limit where the locking mechanism involves nothing beyond the pure phase-restoring dynamics governed by Eq.~(\ref{relativephase}).

To calculate $\Delta S_{2,\mathrm{L}}$, we first need to identify an effective potential energy associated with Eq.~(\ref{relativephase}). Such a potential energy would allow us to calculate the heat released during phase alignment via the first law. Since phase itself does not carry energy, the potential energy is not apparent in Eq.~(\ref{relativephase}). We seek to obtain it by transforming the equation into an energetic form through multiplication by a suitable constant. This is more than a simple rescaling: while the terms without explicit noise scale linearly with the constant, the covariance of the noise term scales quadratically. Thus, the choice of the constant is consequential. We have found $4E_0/\gamma$ to be the appropriate constant. Multiplying Eq.~(\ref{relativephase}) by $4E_0/\gamma$ yields
\begin{equation}
\frac{4E_0}{\gamma}\,d\psi
=
-\frac{\partial U}{\partial\psi}\,dt
+
\sqrt{2\frac{4E_0}{\gamma}k_\mathrm{B}T}\,dW,
\label{thermal}
\end{equation}
where
\begin{equation}
U(\psi)=-\frac{4E_0K}{\gamma}\cos\psi
\label{U0}
\end{equation}
is the sought-after effective potential energy. Equation~(\ref{thermal}) satisfies the fluctuation--dissipation relation, and the von Mises distribution in Eq.~(\ref{vonmises}) takes exactly the Boltzmann form, $p(\psi)\propto\exp[-U(\psi)/(k_\mathrm{B}T)]$, with $U(\psi)$ serving as the potential energy. This agreement with thermodynamic principles provides an operational basis for using $U(\psi)$ as the effective potential energy.

To calculate $\Delta S_{2,\mathrm{L}}$, we increase the normalized locking strength $a'=K'/D$ from zero to $a=K/D$ quasistatically, i.e., slowly enough that the $\psi$ distribution follows the von Mises distribution of  Eq.~(\ref{vonmises}) at all times. The quasistatic work done on the system is $W=\int_0^a \left\langle {\partial U}/{\partial a'}\right\rangle da'=-k_\mathrm{B}T\ln I_0(a)$. The mean potential energy for a given $a'$ is $\langle U\rangle=-k_\mathrm{B}T\,a' I_1(a')/I_0(a')$ and thus $\Delta\langle U\rangle = -k_\mathrm{B}T\,a I_1(a)/I_0(a)$. Therefore, the heat released is $W-\Delta\langle U\rangle$ and $\Delta S_{2,\mathrm{L}} ={(W-\Delta\langle U\rangle)}/{T}$, giving 
\begin{equation}
\Delta S_{2,\mathrm{L}}=k_\mathrm{B}\left[
a\frac{I_1(a)}{I_0(a)}-\ln I_0(a)
\right].
\label{landauer}
\end{equation}
The same result is also readily obtained from Sekimoto's trajectory-level formulation of stochastic heat \cite{Sekimoto}. This $\Delta S_{2,\mathrm{L}}$ during the quasistatic establishment of phase locking exactly cancels the relative-phase entropy reduction $\Delta S_1(\psi)$ in Eq.~(\ref{S1diff}), i.e., $\Delta S_{2,\mathrm{L}}=-\Delta S_1(\psi)$. Thus, this $\Delta S_{2,\mathrm{L}}$ is the Landauer cost of acquiring relative-phase information in the reversible limit \cite{LandauerIBM, Parrondo}. Faster locking produces a larger $\Delta S_{2,\mathrm{L}}$, resulting in $\Delta S_{2,\mathrm{L}}+ \Delta S_1(\psi)>0$. For example, if $a'$ is switched instantaneously from zero to $a$ and $\psi$ is allowed to relax, $\Delta S_{2,\mathrm{L}} + \Delta S_1(\psi)=k_\mathrm{B}\ln I_0(a)>0$. Both quasistatic and fast phase-locking protocols produce the same relative-phase entropy reduction $\Delta S_1(\psi)$, but the latter incurs additional irreversible entropy production.

Once phase alignment is established, the phase-restoring dynamics described by Eq.~(\ref{relativephase}) or Eq.~(\ref{thermal}) with fixed $K$ satisfy detailed balance, with no net heat released into the environment. Thus, in the idealized limit where the locking mechanism involves nothing beyond the pure dynamics described by Eq.~(\ref{relativephase}) or Eq.~(\ref{thermal}), maintaining the already acquired relative-phase information entails no further environmental entropy production by the locking mechanism: $P_\mathrm{L}(\infty)=0$, and hence $\beta=0$. In practical phase-locking circuitry, however, maintaining $K$ even at a constant value generally requires continuous power dissipation and $P_\mathrm{L}(\infty)>0$.

\subsection{Information and thermodynamics of phase alignment}

In the Landauer limit discussed in Sec.~\ref{sec:LL}, the information-theoretic reduction in relative-phase entropy $\Delta S_1(\psi)$, calculated solely from the phase probability density, is matched in magnitude by the environmental entropy production $\Delta S_{2,\mathrm{L}}$ calculated from the energetics of phase alignment using the effective potential energy $U(\psi)$ [Eq.~(\ref{U0})] associated with the pure phase-locking dynamics [Eq.~(\ref{thermal})]. This affirms our treatment of the information-theoretic entropy $S_1$---calculated throughout this paper from the relevant phase probability densities---as a physical entropy. What we see here is the thermodynamics of information \cite{LandauerIBM, Parrondo, Hatano} at work in the establishment of coherence between two oscillators via phase locking. Before concluding this paper, we explore this deep connection between information and thermodynamics further by considering phase locking between oscillators at different frequencies, widely employed in electronic frequency synthesis \cite{LeeBook, RazaviBook, PLL3}. This allows us to examine more closely what constitutes the information acquired upon phase locking and how its physical content is reflected in the thermodynamic cost of acquiring it.

Consider two oscillators operating at different frequencies: our oscillator with frequency $\omega_0$, phase $\phi$, and phase diffusion constant $D$, and a reference oscillator with frequency $\omega_\mathrm{r}$, phase $\phi_\mathrm{r}$, and phase diffusion constant $D_\mathrm{r}$. As before, the reference oscillator is much more precise than our oscillator. Let the two frequencies be related by $\omega_0/N=\omega_\mathrm{r}$ ($N$: positive integer). We consider phase locking that aligns the total phase of our oscillator divided by $N$, $(\omega_0t+\phi)/N$, with the total phase of the reference, $\omega_\mathrm{r}t+\phi_\mathrm{r}$. Phase alignment between $\phi/N$ and $\phi_\mathrm{r}$ is thus established---or the relative phase $\psi\equiv\phi/N-\phi_\mathrm{r}$ is localized around zero---through the locking process. Neglecting the reference noise, the dynamics governing the establishment and maintenance of this phase alignment is given by
\begin{equation}
d\psi=-K\sin\psi\,dt+\sqrt{2(D/N^2)}\,dW,
\label{eq:synthesis_phase}
\end{equation}
which generalizes Eq.~(\ref{relativephase}). The steady-state probability density of $\psi$ in the wrapped representation takes the same von Mises form as in the case of phase locking at identical frequencies, with $a$ now replaced by $a_N\equiv N^2K/D$:
\begin{equation}
p(\psi)=\frac{e^{a_N\cos\psi}}{2\pi I_0(a_N)}.
\label{N-distribution}
\end{equation}
In the strong-locking regime ($a_N\gg1$), $\psi$ is sharply localized around zero, with $\langle\psi^2\rangle\simeq1/a_N$. Thus, $\phi/N$ closely follows the slowly diffusing reference phase $\phi_\mathrm{r}$, and consequently $\phi$ diffuses with a phase diffusion constant $N^2D_\mathrm{r}$ rather than its free-running value $D$  ($N^2 D_\mathrm{r}\ll D$ is typical in practice).

As in Sec.~\ref{sec:LL}, multiplying Eq.~(\ref{eq:synthesis_phase}) by $4E_0N^2/\gamma$ transforms it into an energetic form that satisfies the fluctuation--dissipation relation, with the effective potential energy $U(\psi)=-(4E_0N^2K/\gamma)\cos\psi$. We now increase the locking strength $K'$ quasistatically from zero to $K$, or equivalently, the normalized locking strength $a_N'\equiv N^2K'/D$ from zero to $a_N$. For this process, we calculate the information-theoretic entropy change $\Delta S_1(\psi)$ from the von Mises distribution [Eq.~(\ref{N-distribution})] and the environmental entropy production $\Delta S_{2,\mathrm{L}}$ from the energetics using $U(\psi)$. We obtain
\begin{equation}
-\Delta S_1(\psi)=\Delta S_{2,\mathrm{L}}
=k_{\mathrm B}\left[a_N\frac{I_1(a_N)}{I_0(a_N)}-\ln I_0(a_N)\right].
\label{N-entropy-relation}
\end{equation}
We again reach the Landauer limit, where the information-theoretic entropy reduction is exactly matched by the thermodynamic entropy produced. In the practical strong-locking regime ($a_N\gg1$), Eq.~(\ref{N-entropy-relation}) reduces to
\begin{equation}
-\Delta S_1(\psi)=\Delta S_{2,\mathrm{L}}
\simeq\frac{k_{\mathrm B}}{2}\ln\frac{2\pi a}{e}
+k_{\mathrm B}\ln N.
\label{N-entropy-relation-strong}
\end{equation}

\begin{figure}[t!]
    \centering
    \includegraphics[width=0.65\textwidth]{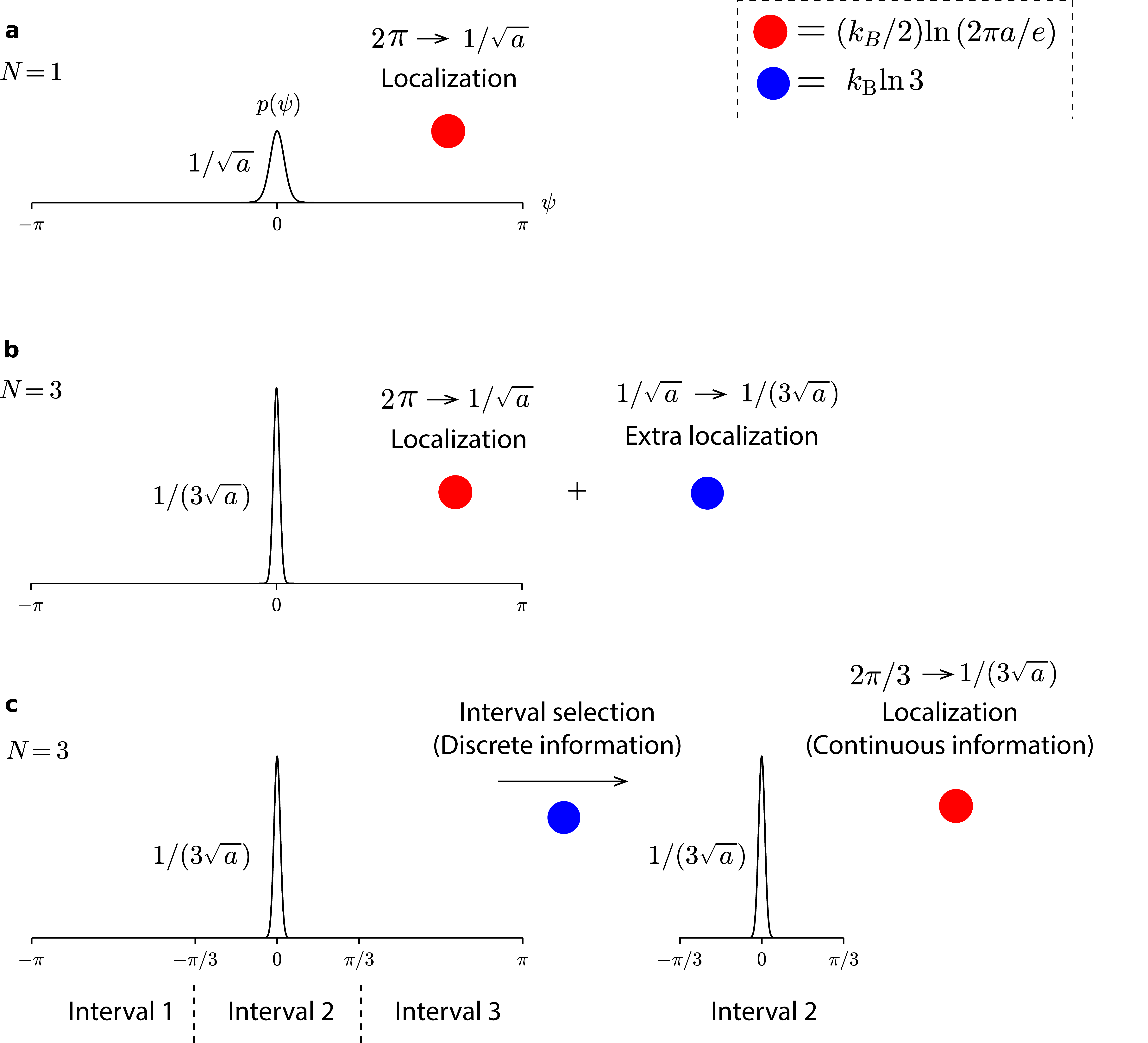}
    \caption{\small Schematic illustration of the acquisition of information about $\psi$ during the establishment of phase locking governed by Eq.~(\ref{eq:synthesis_phase}). (a) $N=1$. (b) $N=3$. (c) An alternative interpretation of the same information gain for $N=3$ in terms of continuous and discrete phase information.}
    \label{fig:information}
\end{figure}

We now examine the information content underlying the correspondence in Eq.~(\ref{N-entropy-relation-strong}) in the strong-locking regime. Phase locking sharply localizes the initially uniform distribution of $\psi$ over the $2\pi$ interval into the von Mises distribution in Eq.~(\ref{N-distribution}) around zero, with a standard deviation of $\sqrt{\langle\psi^2\rangle}\simeq\sqrt{1/a_N}=1/(N\sqrt{a})$, reducing uncertainty about $\psi$ or increasing information about $\psi$. For $N=1$, the information gained by narrowing the distribution from the $2\pi$ interval to a standard deviation of $1/\sqrt{a}$ accounts for the first term in Eq.~(\ref{N-entropy-relation-strong}) (Fig.~\ref{fig:information}a), which is precisely the strong-locking limit of Eq.~(\ref{S1diff}) for phase locking at identical frequencies. For $N >1$, the distribution is narrower by a factor of $N$ relative to the $N=1$ case, yielding an additional information gain of $\log_2 N$ bits or an additional entropy reduction of $k_{\mathrm B}\ln N$, the second term in Eq.~(\ref{N-entropy-relation-strong}) (Fig.~\ref{fig:information}b).
 
There is another way to understand this information gain. Consider $N=3$ as an illustrative example. Before locking, the wrapped phases $\phi$ and $\phi_{\mathrm r}$ are independently and uniformly distributed over the $2\pi$ interval. Now consider $\phi/3$. For any given wrapped value of $\phi$, there are three possible values of $\phi/3$, separated by $2\pi/3$, because phases differing by $2\pi$ are indistinguishable before division but distinguishable after division. Thus, even if we know the wrapped phase $\phi$ exactly, there remains a threefold ambiguity in $\phi/3$. This threefold ambiguity carries over to $\psi=\phi/3-\phi_{\mathrm r}$. We can divide the $2\pi$ interval of $\psi$ into three equal intervals, $\mathrm{Interval~1}=[-\pi,-\pi/3)$, $\mathrm{Interval~2}=[-\pi/3,\pi/3)$, and $\mathrm{Interval~3}=[\pi/3,\pi)$, each containing one of the three possible values of $\psi$ for given wrapped phases $\phi$ and $\phi_{\mathrm r}$. Before locking, these three possibilities are equally probable. We can therefore view the uncertainty in $\psi$ in terms of the discrete uncertainty as to which of the three intervals $\psi$ lies in and the continuous uncertainty as to where within that interval $\psi$ lies.
 
Phase locking, which localizes $\psi$ around zero according to the von Mises distribution with a standard deviation of $1/(3\sqrt{a})$ [Eq.~(\ref{N-distribution}) with $N=3$], reduces both uncertainties (Fig. \ref{fig:information}c). First, since this standard deviation is much smaller than the width $2\pi/3$ of Interval 2, the distribution is effectively confined to Interval 2, resolving the discrete uncertainty as to which of the three intervals $\psi$ lies in. This discrete information gain of $\log_2 3$ bits corresponds to the second term in Eq.~(\ref{N-entropy-relation-strong}) for $N=3$. Second, the same von Mises distribution also provides information about where $\psi$ lies within Interval 2. The ratio of its standard deviation $1/(3\sqrt{a})$ to the interval width $2\pi/3$ is the same as for $N=1$, where the corresponding widths are $1/\sqrt{a}$ and $2\pi$. Thus, the relative degree of localization, and with it the continuous information gain, is the same as for $N=1$. The corresponding entropy reduction must then be the first term in Eq.~(\ref{N-entropy-relation-strong}). In sum, the first and second terms on the right-hand side of Eq.~(\ref{N-entropy-relation-strong}) reflect the acquisition of continuous and discrete phase information, respectively. Remarkably, the equality between $-\Delta S_1(\psi)$ and $\Delta S_{2,\mathrm{L}}$ in the same equation shows that the environmental entropy production---the heat released into the environment divided by $T$---calculated from the energetics of quasistatic phase alignment accounts precisely not only for the continuous information gain but also for the discrete information gain.

The distinction between continuous and discrete phase information becomes even more revealing when we reverse the frequency relation. Consider $\omega_\mathrm{r}=M\omega_0$ ($M$: positive integer), and phase locking that aligns the total phase of our oscillator multiplied by $M$, $M(\omega_0t+\phi)$, with the total phase of the reference, $\omega_\mathrm{r}t+\phi_\mathrm{r}$. Then the relative phase is $\psi=M\phi-\phi_\mathrm{r}$. The analysis above carries over with $N$ replaced by $1/M$, yielding
\begin{equation}
-\Delta S_1(\psi)=\Delta S_{2,L}
\simeq \frac{k_B}{2}\ln\frac{2\pi a}{e}
-k_B\ln M.
\end{equation} 
Unlike phase division, phase multiplication introduces no discrete ambiguity: for a given wrapped value of $\phi$, the wrapped value of $M\phi$ is uniquely determined. The second term thus reflects purely a reduction in continuous information gain by $\log_2 M$ bits, as the standard deviation of the localized distribution of $\psi$, $M/\sqrt{a}$, is larger by a factor of $M$ than in the case of phase locking at identical frequencies.

Finally, combining the two cases, we consider a general rational frequency ratio $\omega_0/\omega_\mathrm{r}=N/M=p/q$, where $p$ and $q$ are coprime positive integers, and $\psi=(q/p)\phi-\phi_\mathrm{r}$ (such fractional frequency relations are widely used for coherence transfer in electronics \cite{LeeBook,RazaviBook}). The foregoing analysis carries over once again with $N$ replaced by $p/q$, yielding
\begin{equation}
-\Delta S_1(\psi)=\Delta S_{2,L}
\simeq
\underbrace{\frac{k_B}{2}\ln\frac{2\pi a}{e}
-k_B\ln q}_{\text{Continuous information}}
+
\underbrace{k_B\ln p}_{\text{Discrete information}}.
\end{equation}
For given wrapped values of $\phi$ and $\phi_\mathrm{r}$, there is a $p$-fold ambiguity in $\psi$. Phase locking resolves this ambiguity, yielding a discrete information gain of $\log_2 p$ bits. Meanwhile, the ratio of the standard deviation $(q/p)/\sqrt{a}$ of the locked $\psi$ distribution to the width $2\pi/p$ of the central interval identified through the resolution of the discrete ambiguity depends only on $q$, resulting in a reduction in continuous information gain by $\log_2 q$ bits relative to the equal-frequency case. Thus, the numerator and denominator of a reduced rational frequency ratio play fundamentally different informational roles. The environmental entropy production calculated from the energetics of quasistatic phase alignment captures precisely these distinct informational roles of the two integers.
 
\section{\normalsize Conclusion}

This paper explored the thermodynamic structure underlying fundamental and indispensable operations of modern electronic systems: the generation and transfer of temporal order using resonator-based self-sustained electronic oscillators. An electronic oscillator is a clock that keeps time by autonomously generating periodic electrical signals. No oscillator is a perfect clock, as thermal noise drives phase diffusion. Greater timekeeping precision---slower phase diffusion---means that an oscillator loses timing information more slowly. Achieving greater precision and delaying the inevitable loss of timing information for longer requires greater environmental entropy production. The resonator quality factor sets the exchange rate between precision and this thermodynamic cost, revealing a distinctive thermodynamic signature of resonator-based self-sustained timekeeping. This thermodynamic cost of maintaining timekeeping precision---the cost of delaying timing information loss---is staggering compared with the amount of timing information the oscillator holds onto before eventually losing it: the typical 1-GHz oscillator in our example produces environmental entropy equivalent to some $10^{17}$ bits to try to hold onto 15 bits of timing information for roughly one second.

Through phase locking, a world-facing oscillator with a lower-quality resonator acquires and retains relative-phase information with a more precise reference oscillator employing a higher-quality resonator (our analysis shows that in the idealized limit, the thermodynamic cost incurred by the phase-locking mechanism in initially acquiring the relative-phase information can reach the Landauer minimum, and reveals how physical information is in the context of phase alignment). The relative-phase information retained in steady state allows the world-facing oscillator to follow the more slowly diffusing reference phase, thereby delaying the loss of its own timing information. Strikingly, delaying timing information loss by retaining relative-phase information is thermodynamically far less costly than delaying timing information loss through autonomous coherence generation. Yet even this greatly reduced thermodynamic cost dwarfs both the amount of relative-phase information retained and the amount of timing information the world-facing oscillator strives to retain but eventually loses. This is the thermodynamic structure associated with the reshaped information content of the phase-locked oscillator pair.

\nocite{apsrev42Control}
\bibliographystyle{apsrev4-2}
\bibliography{references}

\end{document}